\documentclass[12pt,a4paper]{article}
\usepackage{amsmath}
\usepackage[dvips]{graphicx}
\input epsf
\usepackage{times}
\begin{document}
\begin{titlepage}
\title{Antishadowing and multiparticle production}
\author{S. M. Troshin,
 N. E. Tyurin\\[1ex]
\small  \it Institute for High Energy Physics,\\
\small  \it Protvino, Moscow Region, 142280, Russia}
\normalsize
\date{}
\maketitle

\begin{abstract}
We discuss role of absorbtion and antishadowing  in particle production.
Antishadowing, which leads to domination  of
elastic scattering at high energies, appears to
be consistent with  growth of  mean  multiplicity in hadronic collisions.
Moreover, we demonstrate possibility to
reproduce power-like energy behavior of the mean multiplicity in the model
with antishadowing  and discuss  physical implications of such behavior for
the hadron structure.
 \\[2ex]
\end{abstract}
\end{titlepage}
\setcounter{page}{2}

\section*{Introduction}
Multiparticle production and  global  observables such as
mean multiplicity and its energy dependence alongside
with total, elastic and inelastic cross--sections  provide us  a clue
to the mechanisms of confinement and hadronization.
General principles  are very important in the nonperturbative sector,
in particular, unitarity which regulates the relative strength
of elastic and inelastic processes.
Unfortunately,  there are no  universal, generally accepted
methods to implement unitarity
in high energy scattering.  Related problem of absorptive
corrections and their sign
has a long history (cf. \cite{sachbla}) and references therein).

The choice of particular unitarization scheme is not completely a matter
of taste. Long time ago the arguments based on analytical properties of the scattering
amplitude  were put forward \cite{blan} in favor of the rational form of unitarization.
It was shown  that correct analytical properties of the scattering amplitude
in the complex energy plane can be
reproduced much
 easier
in this form of unitarization compared to the most popular
exponential form. Besides that rational form of unitarization
 leads naturally
to  prediction of the
antishadow scattering mode \cite{bdsphl}.  Appearance of this mode
is expected  beyond the Tevatron maximum energy.

Interest in unitarity  and the corresponding limitations
was stimulated under preparation
of the experimental program at the LHC and the future plans
to study soft interactions at the highest energies. Indeed, correct
account for unitarity is also essential under theoretical estimates
of the Higgs production cross-section via the diffractive mechanisms.
The region of the LHC energies is the one where antishadow scattering
 mode is to be presented. It has been demonstrated
 that this mode can be revealed at the LHC directly measuring
 $\sigma_{el}(s)$ and $\sigma_{tot}(s)$ \cite{reltot} and not only through the
 analysis of impact parameter distributions.
 Antishadowing  leads to self--damping of the inelastic channels and
 dominating role of elastic scattering, i. e.
 ${\sigma_{el}(s)}/{\sigma_{tot}(s)}\rightarrow 1$ at $s\to\infty$.
 Natural question arises about consistency of this mechanism with
 the growth with energy of mean multiplicity in hadronic collisions.
Moreover, many models and experimental data suggest power dependence on energy of
mean multiplicity\footnote{Recent discussions of power--like energy dependence
of the mean hadronic multiplicity and list of references to the older papers
can be found in \cite{menon}} and a priori it is not evident
 whether such dependence is compatible with antishadowing or not.

In this note we apply
the rational ($U$--matrix)  unitarization
approach \cite{umat}  for   consideration of the global features
of multiparticle dynamics such as mean multiplicity and role
of absorptive  correction. We show that it is possible to
 reproduce power-like energy behavior of the mean multiplicity in the model
 with antishadowing and discuss its physical implications.

\section{Multiparticle production in the $U$--matrix approach}
The rational form of unitarization
 is based on the relativistic generalization
 of the Heitler equation of radiation dumping \cite{umat}.
 In this approach the elastic scattering amplitude satisfies
unitarity equation since it is a solution  of  the
following equation  \begin{equation} F = U + iUDF
\label{xx} \end{equation}  presented here in the operator form.
 Eq.\ref{xx} allows  one to satisfy unitarity provided the
 inequality \begin{equation} \mbox{Im} U(s,b) \geq 0 \end{equation}
is fulfilled.
The form of the amplitude in the impact parameter representation
is the following:
\begin{equation}
f(s,b)=\frac{U(s,b)}{1-iU(s,b)}, \label{um}
\end{equation}
where $U(s,b)$ is the generalized reaction matrix, which is considered as an
input dynamical quantity similar to the eikonal function.
Analogous form for the scattering amplitude was obtained by Feynman in his
 parton model of diffractive scattering \cite{ravn}.

In the impact parameter representation the unitarity equation
rewritten for the elastic scattering amplitude $f(s,b)$
at high energies has the form
\begin{equation}
\mbox{Im} f(s,b)=|f(s,b)|^2+\eta(s,b) \label{unt}
\end{equation}
where the inelastic overlap function
\[
\eta(s,b)\equiv\frac{1}{4\pi}\frac{d\sigma_{inel}}{db^2}
\]
 is the sum of
all inelastic channel contributions.  It can be expressed as
a sum of $n$--particle production cross--sections at the
given impact parameter
\begin{equation}
\eta(s,b)=\sum_n\sigma_n(s,b),
\end{equation}
where \[\sigma_n(s,b)\equiv \frac{1}{4\pi}\frac{d\sigma_{n}}{db^2},\quad
\sigma_n(s)=8\pi\int_0^\infty bdb \sigma_n(s,b).\]
Inelastic overlap function
is related to $U(s,b)$ as follows
\begin{equation}
\eta(s,b)=\frac{\mbox{Im} U(s,b)}{|1-iU(s,b)|^{2}}\label{uf}.
\end{equation}

Then the unitarity Eq. \ref{unt} points out that the elastic scattering
amplitude at given impact parameter value
is determined by the inelastic processes when the amplitude is
a pure imaginary one.
Eq. \ref{unt} imply the constraint
$|f(s,b)|\leq 1$
 while the ``black disk'' limit
 presumes inequality
$|f(s,b)|\leq 1/2$
and the elastic amplitude satisfying  this condition is a shadow of inelastic
processes. The imaginary
part of the generalized reaction matrix in its turn is the sum of inelastic channel
 contributions:
\begin{equation}
Im U(s,b)=\sum_n \bar{U}_n(s,b),\label{vvv}
\end{equation}
where $n$ runs over all inelastic states and
\begin{equation}
\bar{U}_n(s,b)=\int d\Gamma_n |U_n(s,b,\{\xi_n\}|^2
\end{equation}
and $d\Gamma_n$ is the $n$--particle element of the phase space
volume.
The functions $U_n(s,b,\{\xi_n\})$ are determined by dynamics
 of $h_1+h_2\to X_n$ processes, where $\{\xi_n\}$ stands for the set of respective
 kinematical
 variables.
  Thus, the quantity $\mbox{Im}U(s,b)$ itself
 is a shadow of the inelastic processes.
However, unitarity leads to  self--damping of the inelastic
channels \cite{bbl} and increase of the function $\mbox{Im}U(s,b)$ results in
decrease
 of the inelastic overlap function $\eta(s,b)$ when $\mbox{Im}U(s,b)$ exceeds unity
 (cf. Fig. 1).

\begin{figure}[hbt]
 \vspace*{-0.5cm}
 \begin{center}
 \epsfxsize=120  mm  \epsfbox{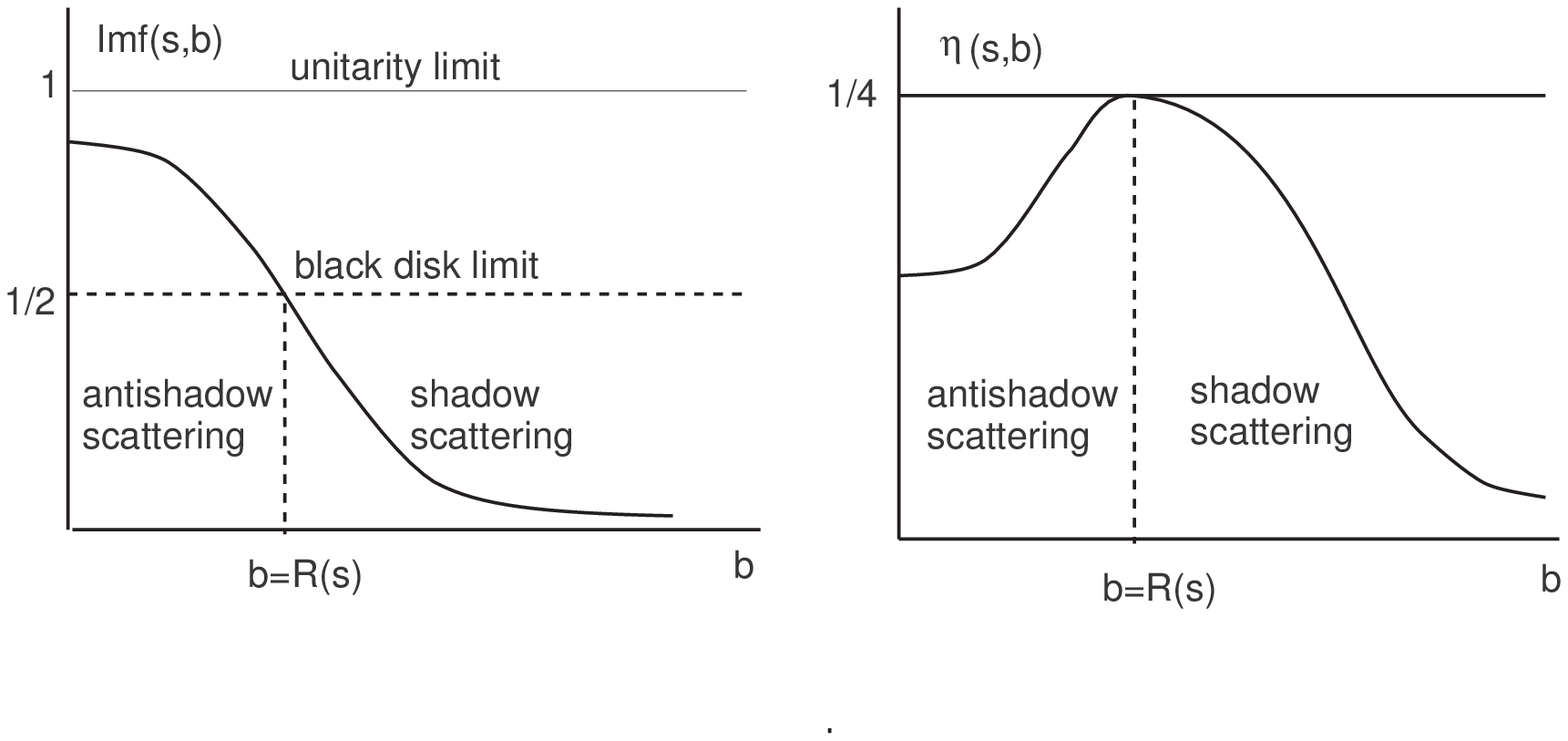}
 \end{center}
 \vspace{-1.5cm}
 \caption{Shadow and antishadow scattering regions}
 \end{figure}
Corresponding inclusive
cross--section
\cite{tmf,gluas} which takes into account unitarity in the direct channel
  has the form \begin{equation}
\frac{d\sigma}{d\xi}= 8\pi\int_0^\infty
bdb\frac{I(s,b,\xi)} {|1-iU(s,b)|^2}.\label{un}
\end{equation}
The function $I(s,b,\xi)$ is expressed via the functions
$U_n (s,b,\xi,\{\xi _{n-1}\})$  determined  by the dynamics
of the processes  $h_1+h_2\rightarrow
h_3+X_{n-1}$:
\begin{equation}\label{idef}
I(s,b,\xi)=\sum_{n\geq 3}n\int d\Gamma_n |U_n (s,b,\xi,\{\xi _{n-1}\})|^2
\end{equation}
and
\begin{equation}\label{sr}
\int I(s,b,\xi)d\xi =\bar n(s,b) \mbox{Im} U(s,b).
\end{equation}

 The kinematical variables $\xi$ ($x$ and $p_\perp$,
 for example) describe the state of the produced particle $h_3$ and
 the set of variables $\{\xi_{n-1}\}$ describe the system $X_{n-1}$
 of $n-1$ particles.

Now we turn to the mean multiplicity and consider first the corresponding
quantity in the impact parameter representation. The $n$--particle production
cross--section $\sigma_n(s,b)$ can be written as
\begin{equation}\label{snb}
\sigma_n(s,b)=\frac{\bar{U}_n(s,b)}{|1-iU(s,b)|^2}
\end{equation}
Then the probability \[P_n(s,b)=\frac{\sigma_n(s,b)}{\sigma_{inel}(s,b)}\] is
\begin{equation}\label{pnb}
  P_n(s,b)=\frac{\bar{U}_n(s,b)}{\mbox{Im} U(s,b)}.
\end{equation}

Thus, we can observe the cancellation of unitarity corrections in
the ratio of cross-sections $\sigma_n(s,b)$ and $\sigma_{inel}(s,b)$.
Therefore the mean multiplicity in the impact parameter representation
\[
\bar n (s,b)=\sum_n nP_n(s,b)
\]
 does not affected by unitarity corrections  and
cannot therefore be proportional
 to $\eta(s,b)$. This conclusion is  consistent with  Eq. (\ref{sr}).
The above mentioned  proportionality is a rather natural assumption in the framework
of the geometrical models, but it is in conflict with
the unitarization. Because of that the results of \cite{enk}
based on such assumption and $U$-matrix unitarization
 should  be taken with precautions.
However, the above cancellation of unitarity corrections
 does not take place for the quantity $\bar n (s)$ which we
 address  in the next section.

\section{Growth of mean multiplicity}

As a starting point
we use a  quark model for the hadron scattering
 described in \cite{csn}.
It is  based on the ideas of chiral quark models.
The picture of a hadron consisting of constituent quarks embedded
 into quark condensate implies that overlapping and interaction of
peripheral clouds   occur at the first stage of hadron interaction (Fig. 2).
Nonlinear field couplings  could transform then the kinetic energy to
internal energy and mechanism of such transformations was discussed
 by Heisenberg \cite{heis} and  Carruthers \cite{carr}.
As a result massive
virtual quarks appear in the overlapping region and  some effective
field is generated.
 Constituent quarks  located in the central part of hadrons are
supposed to scatter in a quasi-independent way by this effective
 field.

\begin{figure}[htb]
\hspace{2.5cm}
\epsfxsize=3in \epsfysize=1.85in\epsffile{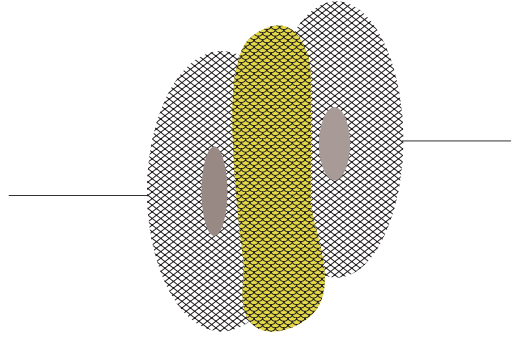}
 \caption[illyi]{Schematic view of initial stage of the hadron
 interaction.}
\label{ill5}
\end{figure}

 Massive virtual quarks play a role
of scatterers for the valence quarks and
 their hadronization leads to the
production of secondary particles.
 To estimate number
of such quarks one could assume that  part of hadron energy carried by
the outer condensate clouds is being released in the overlap region
 to generate massive quarks. Then their number can be estimated  by:
 \begin{equation} \tilde{N}(s,b)\,\propto
\,\frac{(1-\langle k_Q\rangle)\sqrt{s}}{m_Q}\;D^{h_1}_c\otimes D^{h_2}_c,
\label{Nsbt}
\end{equation} where $m_Q$ -- constituent quark mass, $\langle k_Q\rangle $ --
average fraction of
hadron  energy carried  by  the constituent valence quarks. Function $D^h_c$
describes condensate distribution inside the hadron $h$, and $b$ is
an impact parameter of the colliding hadrons.

Thus, $\tilde{N}(s,b)$ quarks appear in addition to $N=n_{h_1}+n_{h_2}$
valence quarks. In elastic scattering those quarks are transient
ones: they are transformed back into the condensates of the final
hadrons. Calculation of elastic scattering amplitude has been performed
in \cite{csn}.

As it was already mentioned  hadronization of massive $\tilde{N}(s,b)$
quarks leads to  formation of the multiparticle
final states, i.e. production of the secondary particles.
Remarkably,  existence of the massive quark-antiquark matter in the stage
preceding
hadronization seems to be
supported  by the experimental data obtained
at CERN SPS and RHIC (see \cite{biro} and references therein).

Since the quarks are constituent, it is natural to expect  a direct
proportionality between the mean multiplicity of the secondary particles in impact
parameter representation and
number of constituent quarks appeared in the collision of the initial hadrons
with  given impact parameter:
\begin{equation}\label{mmult}
\bar n (s,b)=\alpha  \tilde{N}(s,b),
\end{equation}
with a constant factor $\alpha$.
The mean multiplicity $\bar n(s)$ can be calculated according to the
formula
\begin{equation}\label{mm}
\bar n(s)= \frac{\int_0^\infty  \bar n (s,b)\eta(s,b)bdb}{\int_0^\infty \eta(s,b)bdb}.
\end{equation}
It is evident from Eq. (\ref{mm}) and Fig. 1 that the antishadow
mode with the peripheral profile of $\eta(s,b)$ suppress the region of small
impact parameters and main contribution to the mean multiplicity is due to
 peripheral region of $b\sim R(s)$.

To make an explicit calculations  we model for simplicity
 condensate distribution by the exponential form,
  i.e. \[D^{h}_c\sim \exp (-b/R_c).\]
Then we have for the mean multiplicity
\begin{equation}\label{nsbex}
  \bar n (s,b)=\tilde\alpha\frac{(1-\langle k_Q\rangle)\sqrt{s}}{m_Q}\exp (-b/R_c).
\end{equation}
The function $U(s,b)$   is
chosen as a product of the averaged quark amplitudes \begin{equation}
U(s,b) = \prod^{N}_{Q=1} \langle f_Q(s,b)\rangle \end{equation} in
accordance  with assumed quasi-independent  nature  of  valence
quark scattering.
The $b$--dependence of the function $\langle f_Q \rangle$ related to
 the quark formfactor $F_Q(q)$ has a simple form $\langle
f_Q\rangle\propto\exp(-m_Qb/\xi )$.
Thus, the generalized
reaction matrix (in a pure imaginary case) gets
the following  form \cite{csn}
\begin{equation} U(s,b) = ig\left [1+\alpha
\frac{\sqrt{s}}{m_Q}\right]^N \exp(-Mb/\xi ), \label{x}
\end{equation} where $M =\sum^N_{q=1}m_Q$.
At sufficiently high energies where increase of total cross--section
is prominent  we
can neglect the energy independent term  and rewrite
the expression for $U(s,b)$ as  \begin{equation}
U(s,b)=i{g}\left(s/m^2_Q\right)^{N/2}\exp (-Mb/\xi ).
\label{xh} \end{equation}

After calculation of the integrals (\ref{mm})
 we arrive to the power-like dependence
of the mean multiplicity $\bar n(s)$ at high energies
\begin{equation}\label{asm}
\bar n(s) \sim s^\delta,
\end{equation}
where
\[
\delta={\frac{1}{2}\left(1-\frac{\xi }{m_QR_c}\right)}.
\]
\begin{figure}[t]
\hspace{2cm}
\epsfxsize=3.5in \epsfysize=2.5in\epsffile{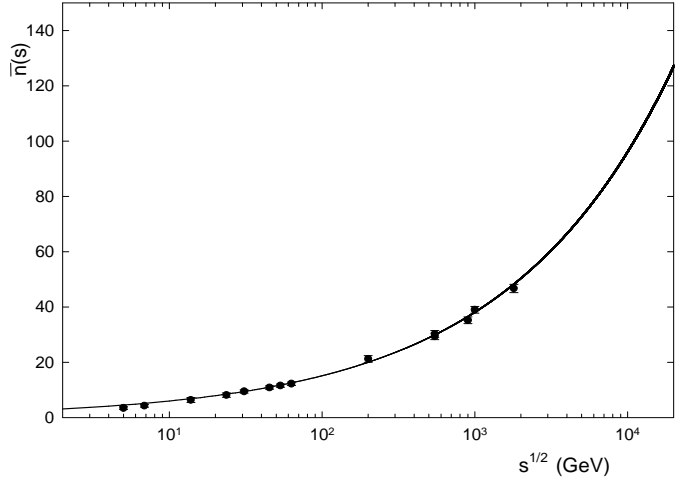}
 \caption[illyi]{Energy dependence of mean multiplicity, theoretical curve
  is given by the equation $\bar n(s)=as^\delta$ ($a=2.328$, $\delta = 0.201$); experimental
  data from \cite{ua5}.}
\label{ill4}
\end{figure}
We have two free parameters in the model, $\tilde\alpha$ and $R_c$, the freedom
of their choice is translated to the free parameters $a$ and $\delta$.
 The value of parameter $\xi=2$ is fixed from the data on angular
 distributions \cite{csn} and for the mass of constituent quark was taken
 the standard value $m_Q=0.35$ GeV. From the comparison
 with experimental data (Fig. 3) on mean multiplicity we obtain that $\delta$
has  value $\delta\simeq 0.2$, which corresponds to effective mass
$M_c=1/R_c\simeq 0.3m_Q$, i.e. $M_c\simeq m_\pi$. It means that condensate
 distribution in the hadron
is rather broad and does not coincide with the distribution of charged matter
given by its formfactor.
The value of mean multiplicity expected at the LHC maximum energy
($\sqrt{s}=14$ TeV) is about 110. Note that the numerical estimates for the
total cross--section and the ratio of elastic to total cross--section
of $pp$--interaction at this energy
 in the model are the following: $\sigma_{tot}\simeq 230$ mb and
${\sigma_{el}(s)}/{\sigma_{tot}(s)}\simeq 0.67$ \cite{reltot}. The latter
value could help to  detect antishadow scattering mode unambiguously.

\section*{Conclusion}
It was shown that the model \cite{csn} based on accounting unitarity and extended to
multiparticle production provides a reasonable description of
the energy dependence of mean multiplicity leading to
its power-like growth with a small exponent. This result
is a combined effect of unitarity and existence of the phase preceding
 hadronization when massive quark--antiquark pairs are generated.
  It is worth noting again that power--like
energy dependence of mean multiplicity appears in various models and is
in good agreement with  heavy--ion experimental data too
\footnote{Recent analysis of mean multiplicity with power--low growth
in Au+Au collisions at RHIC  is given in \cite{bars}}.

Multiplicity distribution $P_n(s,b)$ and mean multiplicity
$\bar n(s,b)$ in the impact parameter representation have
no absorptive corrections, but
antishadowing leads to suppression of particle
production at small impact parameters and the main contribution to
the integral mean multiplicity $\bar n(s)$ comes from
the region of $b\sim R(s)$. Of course, this prediction is valid for
the energy range where antishadow scattering mode starts to develop
(the quantitative analysis of the experimental data
 \cite{pras} gives the value: $\sqrt{s_0}\simeq 2$ TeV)
and is therefore consistent with the ``centrality'' dependence of mean multiplicity
observed at RHIC \cite{phen}.

In addition to the above conclusion, comparison with experimental
data has shown that the peripheral condensate cloud of a hadron
has rather large size.

It is worth also noting  that  no limitations follow from the general principles
of theory for the mean multiplicity, besides the well known one based
on the energy conservation law.
Having in mind relation (\ref{nsbex}), we could say  that the obtained power--like dependence
which takes into account unitarity effects could be considered as a kind of a saturated
upper bound  for the mean multiplicity.

\end{document}